\documentclass[prl,showpacs,preprintnumbers,amsmath,amssymb,twocolumn]{revtex4}

\usepackage{graphicx}
\usepackage{dcolumn}
\usepackage{bm}

\preprint{submitted to Journal of Spintronics and Magnetic
Nanomaterials}

\begin{document}

\title{Magnetic configurations in cubic Bi$_{2}$MnFeO$_6$ alloys from first-principles}

\author{K. Koumpouras}
\author{I. Galanakis}\email{galanakis@upatras.gr}

 \affiliation{Department of Materials Science, School of Natural Sciences,
University of Patras,  GR-26504 Patra, Greece}

\date{\today}

\begin{abstract}
We expand our study on cubic BiFeO$_3$ alloys presented in [K.
Koumpouras and I. Galanakis, \textit{J. Magn. Magn. Mater} 323,
2328 (2011)] to include also the BiMnO$_3$ and Bi$_2$MnFeO$_6$
alloys. For the latter we considered three different cases of
distribution of the Fe-Mn atoms in the lattice and six possible
magnetic configurations. We show that Fe and Mn atoms in all cases
under study retain a large spin magnetic moment, the magnitude of
which exceeds the 3 $\mu_B$. Their electronic and magnetic
properties are similar to the ones in the parent BiMnO$_3$ and
BiFeO$_3$ compounds. Thus oxygen atoms which are the
nearest-neighbors of Fe(Mn) atoms play a crucial role since they
mediate the magnetic interactions between the transition metal
atoms and screen any change in their environment. Finally, we
study the effect of lattice contraction on the magnetic properties
of Bi$_2$MnFeO$_6$.
\end{abstract}


\maketitle

\section{Introduction}

Spintronics brought to the center of the scientific research new
materials with exotic properties \cite{Zutic,Felser,Zabel}. The
latest addition to these materials are the so-called multiferroics
which combine several ferroic orders like ferromagnetism,
ferroelectricity, ferroelasticity etc
\cite{Review,Review5,Review6}. Among them exist some compounds
which combine electric and magnetic order exhibiting the
magnetoelectric effect \cite{Review}. These alloys have several
potential applications like magnetic-field sensors and
electric-write magnetic-read random-access memories
\cite{Hetero,General1,Zhang,Review7,Review8}. Magnetic order and
ferroelectricity have different origins
\cite{Picozzi,Picozzi2,General2} and thus the materials exhibiting
the magnetoelectric effect are few and the coupling between the
magnetic and electric properties is weak. An alternative route to
achieve a strong coupling could be the growth of thin film
heterostructures and several advances have been made towards the
magnetic control of ferroelectricity
\cite{General3,General4,Review2} and the electric control of thin
film magnetism \cite{Films3,Review3,Review4}.

Among the most studied single-component multiferroic compounds are
the bismuth ferrite and the bismuth manganite, BiFeO$_3$ and
BiMnO$_3$. Bulk BiFeO$_3$ crystallizes in  a perovskite-like
pseudocubic structure instead of a ferrite one \cite{Kiselev} and
bulk BiMnO$_3$ prefers a monoclinic lattice \cite{Belik}.
BiFeO$_3$ is a ferroelectric G-type antiferromagnet \cite{Kiselev}
while BiMnO$_3$ is a ferromagnet presenting no net spontaneous
polarization \cite{Baettig,Santos}. A Jahn-Teller distortion
corresponding to an elongation of the oxygen octahedron
surrounding the Mn atom induces a polar order in BiMnO$_3$ and
G-type antiferromagnetism appears as in BiFeO$_3$; this new
multiferroic  state is close in energy to the ferromagnetic
non-polar ground state \cite{Baettig,Santos}. Several
first-principles calculations have been carried out to study the
properties of both bulk BiFeO$_3$ \cite{BiFeO3-Calc} and BiMnO$_3$
compounds \cite{Baettig,Santos,Solovyev}. We refer readers to Ref.
\onlinecite{Palova} for an overview of the literature on both
compounds. Since the single-component crystals like BiFeO$_3$
present only a weak magnetoelectric effect, an alternative route
to achieve a more strong effect has been proposed to be  the
growth of heterostructures where epitaxial strain can enhance the
phenomenon \cite{Films1,Trilayers,Kovachev,Yamauchi}. Doping has
been also proposed to enhance the performance of such structures
\cite{Khomchenko,Cheng}. Towards such heterostructures,
multilayers consisting of alternating layers of BiFeO$_3$ and
BiMnO$_3$ have been proposed where the ferroelectricity of
BiFeO$_3$ couples to the ferromagnetic order in BiMnO$_3$ through
epitaxial strain and a nanoscale checkboard from first-principles
has been recently proposed for several lattice structures
\cite{Palova}. Simple doping of BiFeO$_3$ with Mn leads to an
antiferromagnetic coupling of the Mn spin moments to the Fe ones
\cite{Sosnowska}.

In a recent publication (Ref. \onlinecite{Koumpouras}) we have
presented extended first-principles calculations, employing the
Quantum-ESPRESSO \cite{QE} ab-initio electronic structure method
in conjunction with the Generalized-Gradient Approximation (GGA)
in the Perdew-Burke-Erzenhof formulation \cite{GGA}, on the
electronic and magnetic properties of BiFeO$_3$ alloy as a
function of the lattice constant in the case of the cubic
perovskite structure (see figure 1 in Ref.
\onlinecite{Koumpouras}),  where the heavy cations (Bi) occupy the
corners of the cube, oxygen atoms the center of the faces and the
Fe atom is at the center of the cube. Fe atoms alone form a cubic
cell and the considered 2$\times$2 unit cell in our calculations
contained eight primitive cells and thus eight Fe atoms. We
studied four possible magnetic arrangements: one ferromagnetic
(F-type) and three antiferromagnetic (A-type, C-type and G-type).
We found that all four types of magnetic order are close in energy
and for a lattice constant of the primitive cell larger than 3.888
\AA\ the G-type antiferromagnetism becomes more stable than the
ferromagnetic state. The density of states (DOS) and the spin
magnetic moments of the Fe atoms, which are responsible for the
magnetic properties, showed similar behavior in all possible
magnetic states since magnetic interactions are mediated by oxygen
atoms. Bi atoms showed no contribution to magnetism while the
magnetic properties of the oxygen atoms depended strongly on their
local environment since they are located at the midpoints between
neighboring iron atoms.

\begin{figure}
\includegraphics[width=\columnwidth]{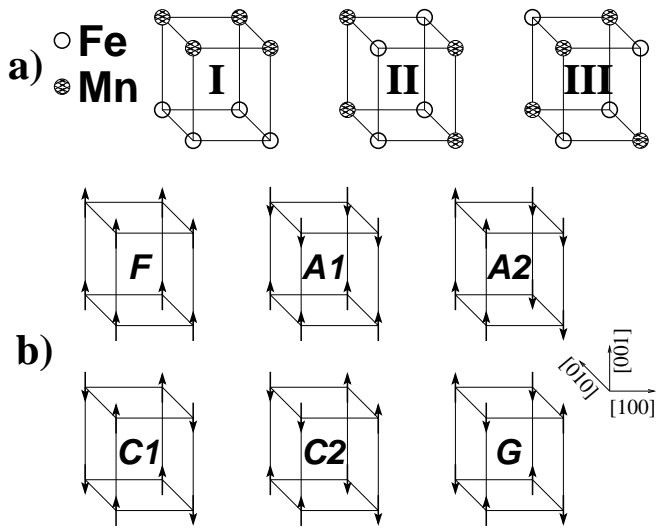}
\caption{Fe and Mn atoms in Bi$_2$MnFeO$_6$ form a cubic lattice.
In (a) we present the three distributions of Mn and Fe atoms which
we considered and in (b) the possible magnetic configurations
(vectors pointing up correspond to positive spin magnetic moments
and vectors pointing down to negative moments). Notice that there
are now two possible A-type antiferromagnetic structures: (i) A1
where atoms belonging to neighboring planes along the [001]
direction have antiparallel spin magnetic moments, and (ii) A2
when this occurs along the [100] directions. For the same reason
there are two C-type antiferromagnetic configurations along the
[110] (C1) and along the [101] (C2) directions. \label{fig1}}
\end{figure}

We expand the study in Ref. \onlinecite{Koumpouras} to cover also
the case of Mn substitution for Fe in BiFeO$_3$ motivated by the
work in Ref. \onlinecite{Palova} since the coexistence of Fe and
Mn cations seems to lead to new multiferroic materials. We have
concentrated our interest on the electronic and magnetic
properties of the resulting compounds employing the same
electronic structure method as in Ref. \onlinecite{Koumpouras} and
calculations details are similar. We have chosen to study a case
with equal number of Fe and Mn atoms (denoted as Bi$_2$MnFeO$_6$)
using the same 2$\times$2 unit cell as in Ref.
\onlinecite{Koumpouras}. Thus now we have four Fe atoms and four
Mn atoms per unit cell. There are several ways to distribute the
Fe and Mn atoms and in panel (a) of Fig. \ref{fig1} we present the
three cases which we have considered. In the I-case we have a
layered structure consisting of alternating pure Fe and pure Mn
layers along the [001] direction. In the II-case we have
alternating pure Fe and pure Mn layers along the [110] direction
and in III-case along the [111] direction. In I-case each Fe(Mn)
atom has four Fe(Mn) and two Mn(Fe) atom as nearest
transition-metal neighbors while in II-case each  Fe(Mn) atom has
two Fe(Mn) and four Mn(Fe) atom as nearest transition-metal
neighbors. In III-case each atom has six neighbors of the other
chemical species. For our calculations we have used a lattice
constant of 3.888 \AA\ for the primitive unit cell in Fig. 1 of
Ref. \onlinecite{Koumpouras}  which is the calculated equilibrium
lattice constant for BiFeO$_3$ in \onlinecite{Koumpouras}.
Moreover due to the lower symmetry there is for Bi$_2$MnFeO$_6$ a
larger number of possible magnetic configurations with respect to
BiFeO$_3$ which are presented in panel (b) of Fig. \ref{fig1}.
Arrows indicate the orientation of the spin magnetic moments of Fe
and Mn atoms at the corners of the cube. F corresponds to the
ferromagnetic configuration. There are two types of
A-antiferromagnetism; in the A1-type all atoms within a (001)
layer are coupled ferromagnetically but successive layers along
the [001] direction are coupled antiferromagnetically between them
while in the A2-type this concerns the layers along the [100]
direction. C1- and C2-types of antiferromagnetism describe
antiferromagnetic coupling of successive layers along the [110]
and [101] directions, respectively. Finally G-type
antiferromagnetism is characterized by alternation of the
orientation of the spin magnetic moments along the [111]
direction. We should note here that for the III-case of the
distribution of Fe and Mn atoms, the symmetry is higher and the
[001] and [100] directions are equivalent and the same stands also
for the [110] and [101] orientations. Thus for this case A1 and A2
as well as C1 and C2 are degenerate. In the next section we
present our calculated results. We have also calculated the case
of BiMnO$_3$ similar to BiFeO$_3$ in Ref.  \onlinecite{Koumpouras}
as reference and at the end of the next section we also present
shortly results for a smaller lattice constant of 3.703 \AA\
corresponding to a contraction of the lattice of about 4.8 \% . In
the last section we conclude and summarize our results.

\begin{table}
\caption{Absolute values of the atom-resolved spin magnetic
moments in $\mu_B$ for all cases and for all magnetic
configurations under study. Results are for a lattice constant of
the primitive unit cell (see Fig. 1 in Ref.
\onlinecite{Koumpouras}) of 14.7 a.u. (3.888 \AA ). In case III
the A1 and A2 structures are equivalent due to symmetry reasons
and the same is valid also for the C1 and C2 configurations. }
\label{table1}
\begin{ruledtabular}
 \begin{tabular}{lcccccccccccc}
 & & &
  \multicolumn{6}{c}{Bi$_2$MnFeO$_6$}\\
&    BiFeO$_3$ &  BiMnO$_3$ & \multicolumn{2}{c}{I-case}  &
\multicolumn{2}{c}{II-case} &
 \multicolumn{2}{c}{III-case}\\
  &  Fe    & Mn  & Fe & Mn & Fe & Mn & Fe & Mn \\
F  & 2.98 & 3.63 & 2.99&3.48 & 3.18 & 3.39 & 3.31 & 3.22\\

A1 & 3.26 & 3.53 & 3.24&3.49 & 3.37 & 3.27& 3.40 & 3.26\\

A2 &      &      & 3.26&3.39 & 3.30 & 3.35 & 3.40 & 3.26 \\

C1 & 3.53 & 3.41 & 3.57&3.32 & 3.51 & 3.37 & 3.59 & 3.31 \\

C2 &      &      & 3.51&3.40 & 3.58 & 3.32 & 3.59 & 3.31\\

G  & 3.65 & 3.24 & 3.68&3.22& 3.68 & 3.21 & 3.68 & 3.18

\end{tabular}
\end{ruledtabular}
\end{table}

\section{Results and discussion}

\begin{figure}
\includegraphics[width=\columnwidth]{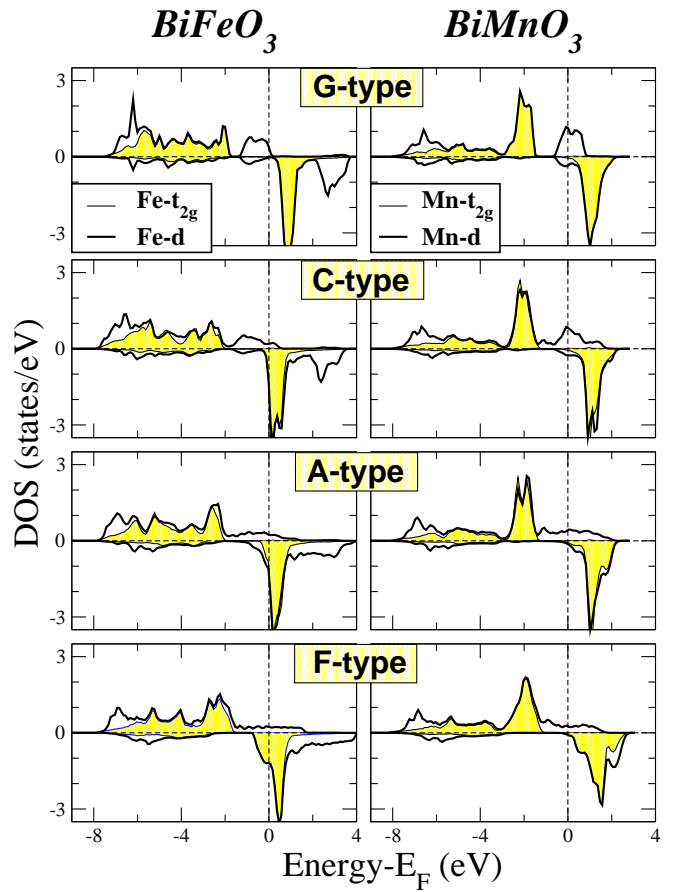}
\caption{(Color online) For both BiFeO$_3$ and BiMnO$_3$ we
present the Fe and Mn density of states (DOS) for all four
possible magnetic configurations. With thick lines is the DOS
projected on the d-orbitals and with the shaded region the
contribution from the triple-degenerated t$_{2g}$ states. The
remaining  contribution comes from the double-degenerated e$_g$
states. Positive DOS values concern the majority-spin electrons
and negative DOS values the minority-spin electrons. The zero
energy value corresponds to the Fermi level.\label{fig2}}
\end{figure}

We will start our discussion from the case of BiMnO$_3$. In Table
\ref{table1} we have gathered the spin magnetic moments in $\mu_B$
for the transition metal atoms for all cases under study and for a
lattice constant of the primitive cell of 3.888\AA . We have also
included in the first column the values for BiFeO$_3$ from Ref.
\onlinecite{Koumpouras}.  Note that for BiMnO$_3$ and BiFeO$_3$
the A1 and A2 as well as the C1 and C2 antiferromagnetic
configurations defined in Fig. \ref{fig1} are degenerated. If we
look at the spin magnetic moments Mn in BiMnO$_3$ posses in
general a very high spin magnetic moment which reaches the 3.63
$\mu_B$ in the ferromagnetic case and varies between 3.24 and 3.53
$\mu_B$ in the antiferromagnetic configurations comparable to the
values for Fe in BiFeO$_3$. The only noticeable difference for the
two alloys concerns the ferromagnetic case (F-type) where Fe has a
spin moment of about 3 $\mu_B$ quite smaller than all other cases.
Mn has one valence electron less than Fe and thus in order to
understand the behavior of its spin magnetic moment with respect
to Fe we have to look at the density of states (DOS) presented in
Fig. \ref{fig2} for Mn and Fe in  BiMnO$_3$ and  BiFeO$_3$,
respectively, for all four magnetic configurations under study. In
the F- and A-configurations the large exchange splitting of the Mn
d-states pushes the minority bands slightly higher in energy with
respect to Fe while for the latter part of the extra electron
occupies minority-spin states near the Fermi level leading to
smaller Fe spin magnetic moments with respect to Mn. In the C- and
G-configurations Mn presents a minority-spin pick pinned exactly
at the Fermi level but this cannot explain the behavior of the
spin magnetic moments in these cases. In the same figure we
present with yellow shaded region the triple degenerated t$_{2g}$
states and thus the rest of the DOS corresponds to the double
degenerated e$_g$ states. The latter are higher in energy with
respect to the t$_{2g}$ states due to the crystal field effect.
The e$_g$ orbitals (d$_{z^2}$ and d$_{x^2+y^2}$) point towards the
neighboring oxygen atoms and thus energetically are unfavorable
with respect to the t$_{2g}$ orbitals (d$_{xy}$, d$_{xz}$ and
d$_{yz}$) which point in the intermediate space, since in the
latter case the Coulomb repulsion with the electrons occupying the
oxygen p-orbitals is smaller. In the minority-spin band of Mn the
e$_g$ are not presented since they are slightly over the upper
bound of the energy axis. In the majority-spin band the e$_g$
states are partially occupied while for Fe the are completely
occupied and thus Mn has a slightly smaller spin magnetic moment
for the C- and G- configurations. Finally we should shortly
discuss the other atoms in these two compounds. Bi atoms are heavy
and their role is to yield the electric polarization when the
lattice is distorted while due to symmetry their spin magnetic
moment is almost zero and thus their contribution to the magnetic
properties can be neglected. Oxygen atoms as discussed in Ref.
\onlinecite{Koumpouras} for BiFeO$_3$ are located at the midpoints
between neighboring transition-metal atoms. If the two Fe atoms
had antiparallel moments the oxygen atom in the middle has a zero
spin magnetic moment due to symmetry, otherwise its spin moment
was about 0.15 $\mu_B$ due to the hybridization between the O-p
and Fe-t$_{2g}$ orbitals. Manganese t$_{2g}$-orbitals hybridize
with the oxygen p-orbitals much weaker and as a result oxygen
atoms in BiMnO$_3$, when both Mn neighboring atoms have parallel
spin moments, exhibit spin magnetic moments of one order of
magnitude smaller than in BiFeO$_3$ (the larger obtained value is
about 0.03 $\mu_B$ in the ferromagnetic configuration). Thus in
BiMnO$_3$ the magnetic properties are entirely localized at the Mn
sites.

\begin{figure}
\includegraphics[width=\columnwidth]{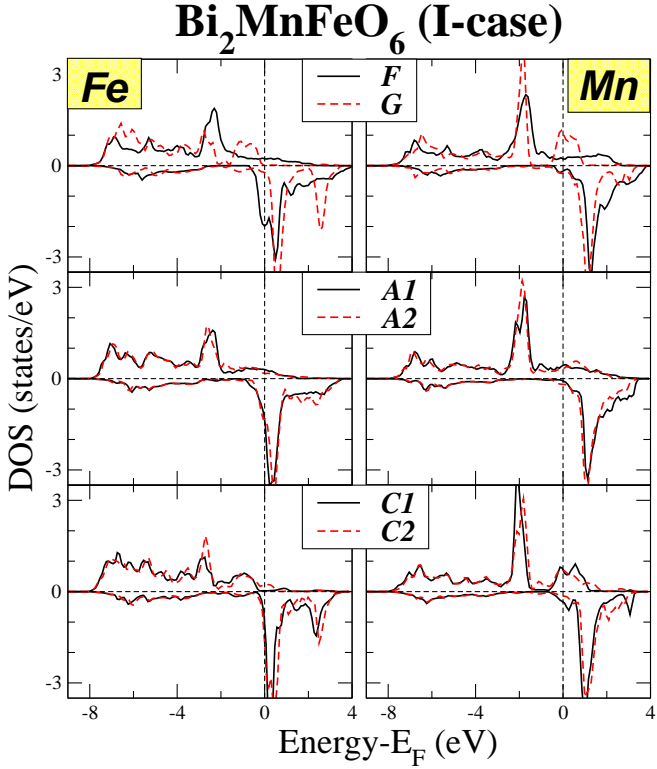}
\caption{(Color online) Fe and Mn d-resolved DOS for the I-case
(see atoms distribution if Fig. \ref{fig1}) for all  six magnetic
configurations. Notice that in all cases the positive DOS values
correspond to the majority states (e.g. in the A1 and A2 cases the
majority states are the spin-up for the Fe atoms and the spin-down
for the Mn atoms). \label{fig3}}
\end{figure}

Probably the most interesting from the three cases of possible
atomic distributions of Fe and Mn atoms presented in Fig.
\ref{fig1} is the I-case where we have along the [001] directions
successive layers of pure Fe and pure Mn atoms. In Table
\ref{table1} we have gathered the spin magnetic moments for all
magnetic configurations under study and in Fig. \ref{fig3} the Fe-
and Mn-resolved DOS. Note in each magnetic configuration all
Fe(Mn) atoms have the same magnitude of the spin magnetic moment
(only its orientation varies) and thus the same DOS shape. We
present the absolute value of the spin magnetic moments and in
Fig. \ref{fig3} the positive DOS values correspond to the
majority-spin electrons in order to make comparison between the
various cases easier to follow. We will start our discussion from
the spin magnetic moments presented in Table \ref{table1}. Our
first remark is that both Fe and Mn spin moments are very close to
their values in the parent BiFeO$_3$ and BiMnO$_3$ alloys,
respectively. Moreover the A1 and A2 as well as the C1 and C2
antiferromagnetic configurations correspond to almost identical
spin magnetic moments. Thus, as discussed in Ref.
\onlinecite{Koumpouras}, we can conclude that the oxygen atoms
play a crucial role. Although themselves posses very small spin
magnetic moments they bridge the transition-metal atoms and they
mediate the magnetic interactions playing a shielding role for the
Fe and Mn atoms. Each Fe(Mn) atom has now 4 Fe(Mn) atoms as
nearest transition-metal neighbors instead of 6 in the perfect
alloys and the other two have been substituted by Mn(Fe) atoms.
But this change in the environment is screened by the intermediate
oxygen atoms and the spin moments of the transition-metal atoms
are only marginally affected. This discussion ia also reflected in
the DOS presented in Fig. \ref{fig3}. For each magnetic
configuration the Fe- and Mn-resolved DOS are almost identical to
the ones for the parent BiFeO$_3$ and BiMnO$_3$ alloys in Fig.
\ref{fig2} and all the details shown in Fig. \ref{fig2} are
present also in the DOS in Fig.  \ref{fig3}; \textit{e.g.} in both
figures Fe in the F-case presents a double pick structure in the
minority-spin band and the Fermi level is pinned exactly between
these two picks while in the G-configuration there two distinct
minority-spin picks corresponding to the t$_{2g}$ and e$_g$
states.

\begin{figure}
\includegraphics[width=\columnwidth]{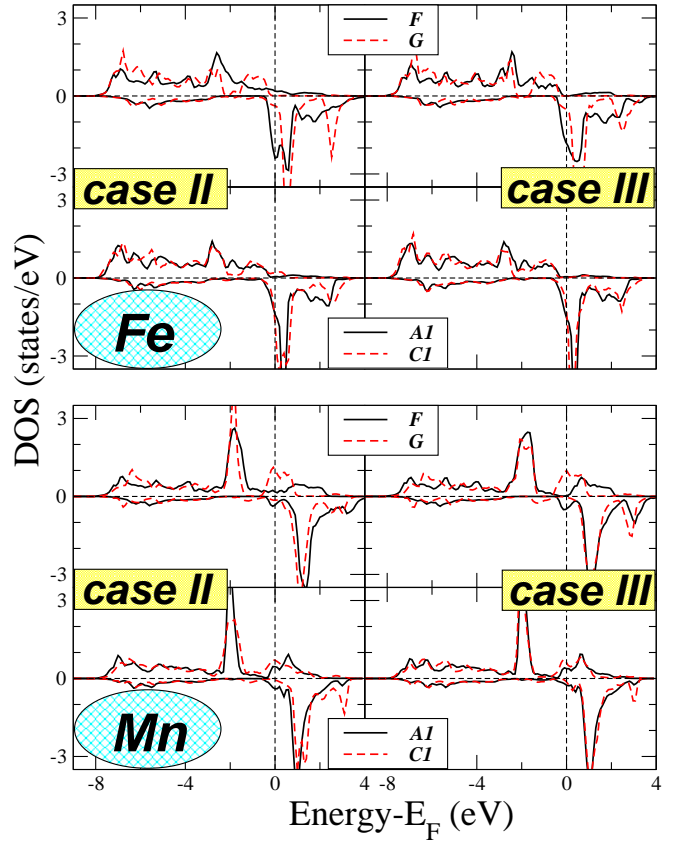}
\caption{(Color online) Same as Fig. \ref{fig3} for Fe (upper
panel) and Mn (lower panel) in Bi$_2$MnFeO$_6$ for the II and III
cases of atoms distribution if Fig. \ref{fig1}. \label{fig4}}
\end{figure}

For the II- and III-cases of the Fe-Mn distribution we include the
spin magnetic moments in Table \ref{table1} and the DOS in Fig.
\ref{fig4}. We do not present the A2 and C2 cases in the figure
since for the II-case they are similar to the A1 and C1 magnetic
configurations while for the III-case as mentioned in the
introduction we cannot distinguish between the A1(C1) and A2(C2)
cases due to the symmetry of the lattice presented in Fig.
\ref{fig1}. In II-case we have [001] chains of pure Fe and pure Mn
or equivalently we can describe the structure as successive layers
of pure Fe and pure Mn along the [110] direction. With respect to
the I-case now each Fe(Mn) atom has only 2 nearest-neighboring
transition metal atoms of the same chemical species instead of 4.
In the III-case, which can be envisaged as alternating layers of
pure Fe and pure Mn along the [111] direction all nearest
neighbors are of the other chemical type, \textit{e.g.} each Fe
atom has 6 nearest-neighboring Mn atoms. Thus although the
magnetic interactions are mediated by the oxygen atoms the effect
on the spin magnetic moments is larger especially for the
ferromagnetic (F-type) alignment of the spin moments. As the
number of Mn neighbors increases the Fe spin magnetic moment in
the F-case increases from 2.99 $\mu_B$ in the I-case to 3.18
$\mu_B$ in the II-case and 3.31  $\mu_B$ in the IIII-case.
Simultaneously the Mn spin magnetic moment drops from 3.48
$\mu_B$ in the I-case to 3.22  $\mu_B$ in the III-case. For the
antiferromagnetic cases under study the influence on the spin
magnetic moments is smaller and for the G-type antiferromagnetism
spin magnetic moments are unaltered by the variation of the Fe-Mn
distribution. Overall even in the F-case the change in the
magnitude of the spin magnetic moments is quite small; less than
0.3  $\mu_B$. The behavior of the spin magnetic moments is also
reflected on the DOS presented in Fig. \ref{fig4}. In the II-case
both Fe- and Mn-resolved DOS for all magnetic configurations under
study are similar to the I-case. In the III case the only
noticeable difference concerns Fe in the ferromagnetic case where
the two picks at the Fermi level in the minority-spin band have
merged in one broad large pick due to the influence of the
exclusively Mn neighbors and this is also the only case where the
variation of the spin-magnetic moment is noticeable reaching the
0.3  $\mu_B$ with respect to the I-case.

Finally, we have also studied the effect of a smaller lattice
constant for the I-case and in Table  \ref{table2} we present the
obtained results for a lattice constant of 3.703 \AA , which is
about 4.8 \%\ smaller than the 3.888 \AA\ used in all previously
presented results. Already in Ref. \onlinecite{Koumpouras} we had
shown that contracting the lattice of BiFeO$_3$ leads to an
important decrease of the spin magnetic moment which in the G-type
antiferromagnetism is less than 1 $\mu_B$ with respect to the 3.65
$\mu_B$ in Table \ref{table1} for the larger lattice constant.
Also in the case of Mn in BiMnO$_3$ the Mn spin magnetic moment
decreases with the contraction of the lattice although the
decrease is not as dramatic as in BiFeO$_3$. This picture is also
reflected in the layered I-case of Bi$_2$MnFeO$_6$ which we
studied. In the I-case the Fe atoms keep a large number of Fe
nearest-neighbors and its spin magnetic moment exhibits an even
larger decrease than in the case of BiFeO$_3$. Especially in the
G-type antifferomagnetism the Fe spin magnetic moments attend a
low of only 0.25 $\mu_B$ being at the edge of being non-magnetic.
On the contrary Mn atoms keep a high value of their spin magnetic
moment which exceeds the 2.8 $\mu_B$ in all cases and thus, as in
the case of BiMnO$_3$, Mn magnetic properties are less affected by
the contraction of the lattice.

\begin{table}
\caption{Same as Table \ref{table1} for a lattice constant of the
primitive cell of 14 a.u. (3.703 \AA ) corresponding to a
contraction of the lattice constant of about 4.8 \% with respect
to the results in Table \ref{table1}. }
 \label{table2}
\begin{ruledtabular}
 \begin{tabular}{lcccc}
&    BiFeO$_3$ &  BiMnO$_3$ & \multicolumn{2}{c}{Bi$_2$MnFeO$_6$ (I-case)} \\
  &  Fe    & Mn  & Fe & Mn \\
F  &2.32 &3.05&2.21&3.08\\

A1 &2.01 &3.03&1.70&3.12 \\

A2 &     &    &1.87&3.05 \\

C1 &1.55&2.93 &1.07&2.81 \\

C2 &    &     &0.76&3.11 \\

G  &0.98&2.53 &0.25&2.84

\end{tabular}
\end{ruledtabular}
\end{table}

\section{Summary and conclusions}

We expand our study in Ref. \onlinecite{Koumpouras} to the case of
Bi$_2$MnFeO$_6$ alloys. First we have studied the BiMnO$_3$ alloy
for various magnetic configurations and we have shown that, as was
the case for BiFeO$_3$, Mn atoms exhibit high values of their spin
magnetic moment, the magnitude of which exceeds the 3 $\mu_B$ in
all cases under study. The small difference between the Fe and Mn
spin magnetic moments can be easily explained by the position of
the e$_g$ states with respect to the t$_{2g}$ states which lie
lower in energy due to the crystal field effect. We took into
account three different possible distributions for the Fe and Mn
atoms in Bi$_2$MnFeO$_6$ and studied one ferromagnetic and five
possible antiferromagnetic alignments of the spin magnetic
moments. Both Fe- and Mn-resolved magnetic and electronic
properties are only marginal sensitive to the various
distributions of the atoms and show properties close to Fe and Mn
atoms in the parent BiFeO$_3$ and BiMnO$_3$ alloys, respectively.
Also among the various magnetic configurations the spin moments
show small variation and their magnitude is  always larger than 3
$\mu_B$. Thus oxygen atoms which are the nearest-neighbors of
Fe(Mn) atoms play a crucial role since they mediate the magnetic
interactions between the transition metal atoms and they screen
any change in their environment. Finally we have shown that
contracting the lattice leads to a large decrease in the spin
magnetic moments of the Fe atoms as in BiFeO$_3$ studied in Ref.
\onlinecite{Koumpouras} while the Mn atoms retain a large portion
of their spin magnetic moment.

%---------------------------------------------------------------------------------------------------------------------------

Authors acknowledge financial support from the K. Karatheodori
program Nr. C588 of the University of Patras.

\end{document}